\documentclass[preprint,showpacs,amsmath,amssymb]{revtex4}
\usepackage{graphicx}
\usepackage{dcolumn}% Align table columns on decimal point
\usepackage{bm}% bold math

\begin{document}

\title{ Baryon magnetic moments  in colored quark cluster model}

\author{Qing-Wu Wang$^{1,3}$, Xi-Guo Lee$^{1,2}${\footnote{Email: xgl@impcas.ac.cn}}, Shi-Zhong Hou$^{1,3}$, and Yuan-Jun Song$^{4}$ }

\affiliation{$^{1}$Institute of Modern Physics, Chinese Academy of
Science, P.O. Box 31, Lanzhou 730000, P.R.China}
\affiliation{$^{2}$Center of Theoretical Nuclear Physics, National
Laboratory of Heavy Ion Collisions, Lanzhou 730000, P.R.China}
\affiliation{$^{3}$Graduate School, Chinese Academy of Science,
Beijing 100049, P.R.China} \affiliation{$^{4}$Department of
Physics, Hebei North University, Zhangjiakou, 075000,P.R.China}
%\date{\today}

\begin{abstract}

Using the colored quark cluster model, we study magnetic moments
of the octet baryons. We give the values of the magnetic moments
of baryons $p$, $n$, $\Sigma^+$, $\Sigma^-$, $\Xi^0$, and $\Xi^-$.
The results also show that the orbital motion has very significant
effects on the spin and magnetic moments of those baryons and the
strange component in the proton is very small.

\end{abstract}
 \maketitle{}

\section{INTRODUCTION}

 It is clear now that the quark sea in nucleon has a nontrivial
 contribution. The EMC effects have indicated that only a
small amount of nucleon spin is carried by quark, and  that the
strange sea quark in the proton may have a negative contribution
to the nucleon spin\cite{spin,Bass,Jaffe}. Recently, both deep
inelastic scattering and Drell-Yan  experiments show that  there
exists flavor asymmetry of light quarks in the nucleon
sea\cite{drellyan,Garvey}. Moreover, the parity-violating
electron-nucleon experiments indicate that the strangeness
electric form factor $ G_E^s (q^2) $ is negative while the
strangeness magnetic form factor $ G_M^s (q^2)$ is
positive\cite{Spayde,Aniol,Mass,Armstrong}.  These issues mean
that the constituent valence quark model(CVQM) cannot explain the
complicated quark structure of baryons completely.

Beyond the CVQM, more interesting models about the substructure of
baryons have been proposed. Such as quark-gluon hybrid model,
diquark-quark model \cite{diquark}, various meson cloud
model\cite{mesoncloud,brodsky,klambda,Chen}, and so on. To
investigate the possible strange component in nucleon, there have
many theoretical approaches, such as the lattice QCD calculation
\cite{lattice}, chiral perturbation and dispersion
relation\cite{chiral}, GDP sum rule\cite{GDP}, $K^+ \Lambda$ meson
cloud model\cite{klambda}, and various quark models\cite{quark}
correlating the octet baryon magnetic moments by assuming SU(3)
flavor symmetry\cite{su3,karl}. Most of the theoretical analyses
and calculations have given a negative sign for magnetic form
factor of proton. Ref.\cite{Hong} has obtained a positive result,
however, the positive contribution is believed being automatically
included by a relativistic calculation\cite{Chen}.

More recently, a new colored quark cluster model (CQCM) has been
proposed. In this model, the $qqqq\bar q$ fluctuation tends to
arrange itself into energetically more favorable states. Of the
strange components, there is a unique $uuds\bar s$ configuration
which can give the right signs of the strange magnetic, electric
and axial form factors\cite{zou,riska}. In this configuration, the
$\bar s$ is in the ground state and the $uuds$ subsystem in the
$P$-state. This configuration has the lowest energy of all
configurations under the assumption that the hyperfine interaction
between the quarks is spin dependent\cite{helminen}. This
configuration may also give an explanation to the excess of $\bar
d$ over $\bar u$\cite{Garvey}. Besides, the five-quark components
of this configuration in the $\Delta(1232)$ give a significant
contribution  to $\Delta(1232)\to N\pi$ decay
 \cite{deltadecay}.

The purpose of this paper is to study the sea  quark contributions
to baryon magnetic moments. In Sec.II, we obtain a general formula
of the magnetic moments of octet baryons by using the CQCM. In
Sec.III we give our numerical results by fitting to experimental
values of octet baryon magnetic moments. The results show that the
theoretical values of those magnetic moments from the CQCM is
better than those from the CVQM, and that the strange component in
the nucleon is small or zero. In Sec.IV, we have compared our
results with the already existed experiments and other theoretical
analyses.

\section{THE BARYON MAGNETIC MOMENT}

We now discuss the magnetic moment of octet baryons. In the CQCM
of Ref.\cite{zou}, the positive parity demands that the four-quark
subsystem is orbitally excited to P-shell with a spatial symmetry
$[31]_X $. In order to give a colored singlet state the four-quark
subsystem has the color state $[211]_C $ since the anti-quark is
in the $[11]_C$ representation. The Ref.\cite{zou} also indicates
that configuration $[4]_{FS} [22]_F [22]_S $ is outstanding for
its energy is some 140-200MeV lower than any other configuration
if the hyperfine interaction between the quarks is described both
by the flavor and spin dependent hyperfine interaction
$-C\Sigma_{i<j}\vec \lambda {}_i \cdot \vec \lambda {}_j\vec
\sigma {}_i \cdot \vec \sigma _j
 $, where $C$ is a constant with the value $\sim $20-30MeV. This
 hyperfine interaction has led to the empirical ordering of the baryon resonances\cite{helminen}.
 The total wave function of the 5-q state with spin +1/2 is
written as

\[\left| {B, + \frac{1}{2}} \right\rangle  = A_5 \sum\limits_{abcde}
{\sum\limits_{Ms_z^\prime ms_z } {C_{JM,\frac{1}{2}s_z^\prime
}^{\frac{1}{2}\frac{1}{2}} C_{1m,Ss_z }^{JM} C_{[31]_a [211]_a
}^{[1^4 ]} } } C_{[31]_b [FS]_c }^{[31]_a } C_{[F]_d [S]_e
}^{[FS]_c }\]
\begin{equation} \times[31]_{X,m} (b)[F]_d [S]_{s_z } (e)[211]_C (a)\bar \chi
_{s_z^\prime } \varphi (\{ r_i \} ).
  \label{eq:wave}
  \end{equation}
Here we use Weyl tableaux to represent the flavor, spin and color
state wave function\cite{chenjinquan}.  The capital $C$ with
superscripts and subscripts denotes the Clebsch-Gordan(CG)
coefficient. The $\bar \chi _ {s_z^\prime}$ is the spin state of
the anti-quark and $\varphi (\{ r_i \} )$
 a symmetric function of the coordinates of
the 5-q system. $A_5$  denotes the amplitude of the 5-q component.
For the mixed flavor symmetry representation $[22]_F$ of the
$uuds$ system, two independent flavor wave functions are written
as
 \[\left| {[22]_{F1} } \right\rangle  = \frac{1}{{\sqrt {24} }}[2\left| {uuds} \right\rangle  + 2\left| {uusd} \right\rangle  + 2\left| {dsuu} \right\rangle  + 2\left| {sduu} \right\rangle  \]
  \[- \left| {duus} \right\rangle  - \left| {udus} \right\rangle  - \left| {sudu} \right\rangle  - \left| {usdu} \right\rangle \]
\begin{equation} - \left| {suud} \right\rangle  - \left| {dusu} \right\rangle  - \left| {usud} \right\rangle  - \left| {udsu} \right\rangle
],
 \end{equation}
\[ \left| {[22]_{F2} } \right\rangle  = \frac{1}{{\sqrt 8 }}[\left| {udus} \right\rangle  + \left| {sudu} \right\rangle  + \left| {dusu} \right\rangle  + \left| {usud} \right\rangle
\]
 \begin{equation} - \left| {duus} \right\rangle  - \left| {usdu} \right\rangle  - \left| {suud} \right\rangle  - \left| {udsu} \right\rangle
 ].
 \end{equation}
And the two spin functions of $[22]_S$  can be obtained by the
substitutions $u \longleftrightarrow \uparrow $ and $d,s
\longleftrightarrow \downarrow$ with a proper normalization
factor. Using $0$ and $1$ to denote the ground-state and $P$-state
wave functions for the constituent quarks, the spatial wave
functions of $[31]_X$ are
\begin{equation}
\left| {[31]_{X1} } \right\rangle  = \frac{1}{{\sqrt {12}
}}[3\left| {0001} \right\rangle  - \left| {0010} \right\rangle  -
\left| {0100} \right\rangle  - \left| {1000} \right\rangle ],
 \end{equation}
\begin{equation}
\left| {[31]_{X2} } \right\rangle  = \frac{1}{{\sqrt 6 }}[2\left|
{0010} \right\rangle  - \left| {0100} \right\rangle  - \left|
{1000} \right\rangle ],
 \end{equation}

 \begin{equation}
\left| {[31]_{X3} } \right\rangle  = \frac{1}{{\sqrt 2 }}[\left|
{0100} \right\rangle  - \left| {1000} \right\rangle ].
  \end{equation}

From Ref.\cite{zou} we know that the $[4]_{FS} [22]_F [22]_S $
configuration does not allow $uudu\bar u$ in the proton. This is
consistent with the observed excess of $\bar d$  over $\bar u$
\cite{Garvey}. The proton may also have an admixture with the
flavor-spin symmetry $[4]_{FS} [31]_F [31]_S $, in which case no
suppression exists. However, it is energetically less favorable.
The empirical evidence  for the large flavor asymmetry of the $q
\bar q$ components\cite{Garvey} suggests that this configuration
$[4]_{FS} [31]_F [31]_S $ should have a smaller probability than
the favored one $[4]_{FS} [22]_F [22]_S $. We do not consider the
meson cloud contribution here, and the quark wave function for
proton may now be expressed as
   \begin{equation}
   \left| p \right\rangle  = A_3 \left| {uud} \right\rangle  + A_{5d}
   \left| {uudd\bar d} \right\rangle  + A_{5s} \left| {uuds\bar s} \right\rangle
  \end{equation}
with the normalization condition  $\left| {A_3 } \right|^2  +
\left| {A_{5d} } \right|^2  + \left| {A_{5s} } \right|^2  = 1$.

The nonperturbative sea quark effects have also been studied in
baryons other than nucleons\cite{susumu}. Taking the same
consideration above for proton, we can write the wave functions of
baryons $p,n,\Sigma^+,\Sigma^-,\Xi^0$, and $\Xi^-$ in a general
form as

\begin{equation}
   \left|B \right\rangle  = A_3 \left| {\alpha\alpha\beta} \right\rangle  + A_{5\beta}
   \left| {\alpha\alpha\beta\beta\bar \beta} \right\rangle  + A_{5\gamma} \left| {\alpha\alpha\beta\gamma\bar \gamma}
   \right\rangle,\label{eq:waveB}
  \end{equation}
where the $\alpha,\beta$, and $\gamma$ can be taken as $d, u$, and
$s$ quarks for neutron, $u,s$, and $d$ for $\Sigma^+$, etc. The
5-q components only contain the configuration of $[4]_{FS} [22]_F
[22]_S $ here and we use this representation throughout this
paper. Because the $\Sigma^0$ and $\Lambda^0$ have three different
valence quarks, we do not consider them here.

In the non-relativistic quark model, the magnetic moment
contribution of quark to the proton magnetic moment is defined as
the expectation value of the following operator
\begin{equation}
\begin{array}{*{20}c}
   {\hat \mu _i  = \frac{{\hat Q_i }}{{2m_i }}(\hat l_i  + \hat \sigma _i ),} & {} & {i = u,d,s.}  \\
\end{array}
\end{equation}
Here  $\hat Q_i $ is the electrical charge operator, and  $m_i $
 the constituent quark mass. In the  naive quark model the
proton consisting of two $u$ quarks and one $d$ quark, all in a
relative S-wave. Using the $ SU_6^{\sigma f}  \supset SU_2^\sigma
\times SU_3^f $ symmetrical wave function as an approximation, the
magnetic moment contribution of the $uud$ component  in unit of
nuclear magneton (n.m) is
\begin{equation}
\left\langle {uud} \right|\sum\limits_i {\hat \mu _i } \left|
{uud} \right\rangle={4 \over 3}{e_u}{{m_p } \over {m_u }}    - {1
\over 3}{e_d}{{m_p } \over {m_d }},
\end{equation}
where the $e_u$ and $e_d$ denote the electric charges of $u$ quark
and $d$ quark respectively, and in the following $e_q$($e_{\bar
q}$) the corresponding quark(anti-quark) electric charge. $m_u$
and $m_d$ are the $u$ and $d$ quark mass, while $m_p$ the mass of
proton. And for baryon $B = \alpha \alpha \beta $\cite{Brekke},
\begin{equation}
\left\langle {\alpha\alpha\beta} \right|\sum\limits_i {\hat \mu _i
} \left| {\alpha\alpha\beta} \right\rangle={4 \over
3}{e_\alpha}{{m_p } \over {m_{\alpha} }} - {1 \over
3}{e_\beta}{{m_p } \over {m_{\beta} }}.
\end{equation}

 Within the 5-q components, the 4-q subsystem gives
no spin contribution to the magnetic moment because symmetry
$[22]_s $ gives spin zero. But every quark in this subsystem has
probability of being excited to P-state, which gives an orbital
magnetic moment
 \begin{equation}
\mu ^{(l)}_q = \left\langle {qqqq\bar q, + \frac{1}{2}}
\right|\frac{{\hat Q_q }}{{2m_q }}\hat l\left| {qqqq\bar q, +
\frac{1}{2}} \right\rangle={ e_q \over 6}{{m_p } \over {m_q
}}P_{5q}.
\end{equation}
 Here $P_{5q}  = \left| {A_{5q} } \right|^2 $ is the
probability of 5-q components. The anti-quark in its ground state
gives a magnetic contribution
 \begin{equation}
 \mu _{\bar q} =
\left\langle {qqqq\bar q, + \frac{1}{2}} \right|\frac{{\hat Q_q
}}{{2m_{\bar q} }}\hat \sigma \left| {qqqq\bar q, + \frac{1}{2}}
\right\rangle=-{e_{\bar q} \over 3}{{m_p } \over {m_{\bar q}
}}P_{5q}.
\end{equation}

Besides these diagonal contributions from quark spin and orbital
motion, the transitions or non-diagonal matrix elements between
the $qqqq\bar q$ and $qqq$ components may have some contributions,
too. However, these non-diagonal contributions depend both on the
explicit wave function model and the model for $q\bar q \to
\gamma$ vertices. Cases will be more complicated to take into
account the confining interaction between the quarks that leads to
bound state wave functions\cite{riska}. With a lot of unexplicit
parameters, the results will be very ambiguous. So, for
simplicity, we neglect the contributions of these transition
matrix elements in this paper. Then, by adding all the spin and
orbital angular contributions to the magnetic moment, the total
magnetic moment of polarized proton is
\begin{equation}
\mu _p = P_{3}({4 \over 3}{e_u}{{m_p } \over {m_u }}    - {1 \over
3}{e_d}{{m_p } \over {m_d }})+ P_{5d} (\sum\limits_{q = u,u,d,d}
{e_q \over 6}{{m_p } \over {m_q }}  - {e_{\bar d} \over 3}{{m_p }
\over {m_{\bar d }}}) + P_{5s} (\sum\limits_{q = u,u,d,s} {{e_q
\over 6}{{m_p } \over {m_q }}}  - {e_{\bar s} \over 3}{{m_p }
\over {m_{\bar s }}}). \label{eq:moment}
\end{equation}

From Eq.\eqref{eq:waveB}, we obtain the general form of the
magnetic moment for the six baryons as
\begin{equation}
\mu _B = P_{3}({4 \over 3}{e_\alpha}{{m_p } \over {m_{\alpha} }} -
{1 \over 3}{e_\beta}{{m_p } \over {m_{\beta} }}) + P_{5\beta}
(\sum\limits_{q = \alpha,\alpha,\beta,\beta} {{e_q \over 6}{{m_p }
\over {m_q }}}  - {e_{\bar \beta} \over 3}{{m_p } \over {m_{\bar
\beta }}}) + P_{5\gamma} (\sum\limits_{q =
\alpha,\alpha,\beta,\gamma} {{e_q \over 6}{{m_p } \over {m_q }}} -
{e_{\bar \gamma} \over 3}{{m_p } \over {m_{\bar \gamma }}}),
\label{eq:moment2}
\end{equation}
with the normalization condition $P_{3}+P_{5
\beta}+P_{5\gamma}=1$.
\bigskip
\section{THE GLOBAL FIT AND RESULTS}

The Eq.\eqref{eq:moment2} means that the baryon magnetic moments
depend on the quark masses and the probabilities of those 5-q
components. To reduce the number of these parameters, we assume
that those $P_3$ are equal for the six baryons, ie,
$P^{p}_3=P^{n}_3=P^{\Sigma^+}_3=P^{\Sigma^-}_3=P^{\Xi^0}_3=P^{\Xi^-}_3$,
and we hold the same assumption for $P_{5\beta}$ and
$P_{5\gamma}$. As in Ref.\cite{karl}, we use a fit method to
discuss these parameters. In order to reduce these parameters
further, we take $m_u=m_d$, $m_q=m_{\bar q}$. As a result, we see
that the six baryon magnetic moments from Eq.\eqref{eq:moment2}
contain only four parameters now, ie, $m_u$, $m_s$, $P_{5\beta}$,
and $P_{5\gamma}$. To give the concrete values of these
parameters, we consider the relatively simple but commonly used
method, namely, to minimize the following function\cite{karl}:
\begin{equation}
\chi ^2  = \sum\limits_{k = 1}^m {\frac{{(T_k  - E_k )^2
}}{{\sigma _k^2 }}}, \label{eq:chi}
\end{equation}
where $E_k $ is the measured value, and $T_k $ the corresponding
theoretical value. m, the number of the baryons, is six here. The
error $\sigma^2 _k $ is taken to be the addition of  a theoretical
error and experimental error in quadrature  as in Ref.\cite{karl}.
The theoretical error comes from a comparison of the sum rule
\begin{equation}
\mu (n) - \mu (p) + \mu (\Sigma ^ +  ) - \mu (\Xi ^0 ) + \mu (\Xi
^ -  ) - \mu (\Sigma ^ -  ) = 0
\end{equation}
with experimental data. The left hand of this equation is actually
$ - 0.49 \pm 0.03\ n.m$. If the errors are equally shared among
the six baryons, the theoretical error may be  $0.49/6 \sim 0.08 \
n.m$. In Table \ref{table1}, we list the experimental data from
PDG\cite{pdg}.

 we firstly examine the value of $P_{5 \gamma}$ given in
Ref.\cite{riska}. In that work, an analysis for the preferred
configuration $[4]_{FS} [22]_F [22]_S $ shows that the qualitative
features of empirical strangeness form factors may be described
with a $ ~15\% $ admixture of   $uuds\bar s$ in the proton. Fixing
the $P_{5\gamma}$ with this value and taking only the $P_{5\beta}$
as variable, the minimum of  $\chi ^2 $ happens at
$P_{5\beta}=0.13$, ie, the probability of $uudd\bar d$ in proton
is 0.13. We express the minimum of $\chi ^2 $ as $\chi _m ^2 $.
The fitting result $\chi_m ^2  = 18.07$ is not better than that
from the CVQM.  The fitting results of the CVQM
$(P_{5\alpha}=P_{5\beta}=0)$ are showed in the Table \ref{table1}
with quark masses being $m_u=344.03$MeV and $m_s=544.76$MeV.

Then, we take both the probabilities of 5-q components
$P_{5\beta}$ and $P_{5 \gamma}$ as variables to minimize the
Eq.\eqref{eq:chi}. And we find that there are two areas where the
minimum can occur: the 3-q component or the 5-q component dominant
in the baryons. As we know, however, the CVQM has success in
low-lying baryon spectroscopy and magnetic moment. The
probabilities of non-perturbative sea-quarks may not be very
large. So, we only consider the case that the probabilities of 5-q
components are small.

\begin{figure}
 \begin{minipage}[t]{.45\linewidth}
\begin{center}
\fbox{
\includegraphics[height=6cm,width=8cm]{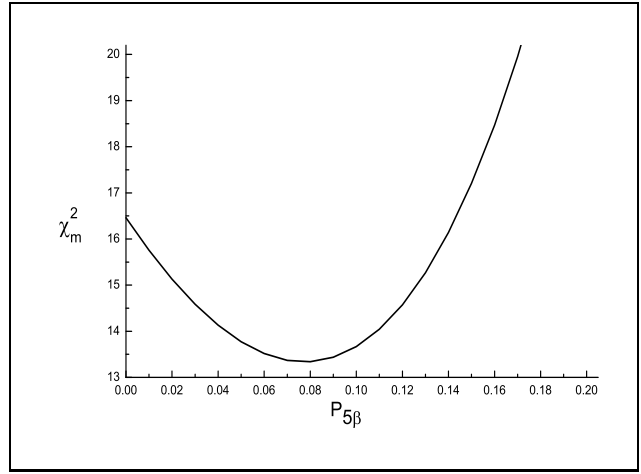}}

\caption{The minimum of quantity $\chi^2$  vs the probability of
$\alpha \alpha \alpha \beta \bar \beta$  in case of
$P_{5\gamma}=0$. When $P_{5\beta}>0.14$, the $\chi_m^2$ will be
large than the value from the naive quark model, which corresponds
to  the  point of $P_{5\beta}=0$.}

\label{fig1}
\end{center}
 \end{minipage}
\end{figure}

\begin{figure}[h]
\begin{minipage}[t]{.45\linewidth}
\begin{center}
\fbox{
\includegraphics[height=6cm,width=8cm]{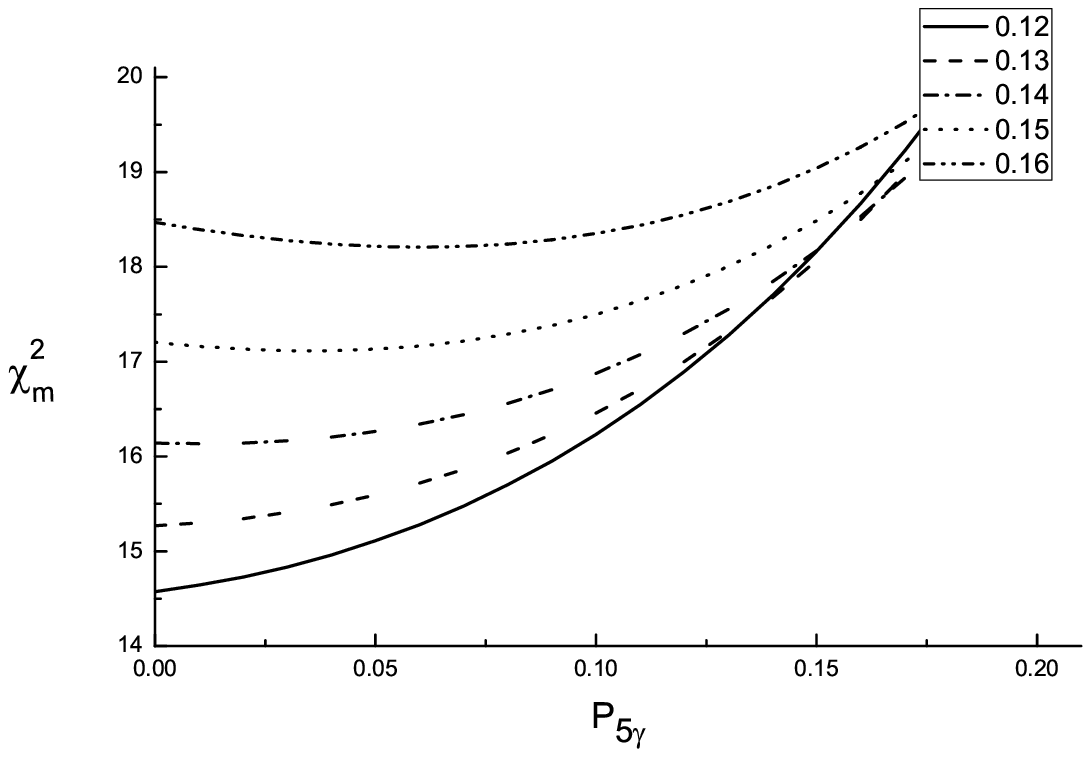}}

\caption{The $\chi _m^2$ as a function of the probability of
$\alpha\alpha\beta\gamma\bar \gamma$ component at several fixed
points of $P_{5\beta}$. Only the $P_{5\beta}$ is less than $0.14$
and the $P_{5\gamma}$ is less than $0.10$, can make the $\chi
_m^2$ be small than the value from naive quark model. }

\label{fig2}
\end{center}
 \end{minipage}
\end{figure}

  The mathematical minimum of $\chi$ is $\chi _m ^2 = 13.36$ with
$P_{5\gamma}=0$ and $P_{5\beta}=0.08$, which is inconsistent with
the observed excess of $\bar d$ over $\bar u$ in proton with $\bar
d - \bar u = 0.12$ \cite{Garvey}. Considering this excess, the
best fitting result is  $\chi_m ^2  = 14.57$, with
$P_{5\gamma}=0$, $P_{5\beta}=0.12$. In this case, we obtain the
magnetic moments of baryons labelled as CQCM in
Table.\ref{table1}. The corresponding quark masses are $
m_u=300.84$MeV and $m_s=463.74$MeV. The $\chi_m ^2 $ as a function
of $P_{5\beta}$ with $P_{5\gamma}=0$ is presented on
Fig.\ref{fig1}. And we find, if $P_{5\gamma} $ is not zero and to
ensure the $\chi_m ^2 $ less than that deduced from the naive
quark model, the value of $P_{5\gamma} $ needs to be less than $
10\% $. This means that the the probability of strange component
in the proton cannot be more than $ 10\% $. On Fig.\ref{fig2}, we
plot $\chi_m^2$ as function of $P_{5\gamma} $ at some fixed points
of $P_{5\beta} $. Both form Fig.\ref{fig1} and Fig.\ref{fig2} we
can see that the up limit of the probability $P_{5\gamma} $ is
about $14\%$, ie, the probability of the $uudd\bar d$ may not be
more than $14\%$ in proton.

In Table \ref{table1}, the theoretical errors are computed as
$(E-T)/E$, where $E$ is the experimental magnetic moment value,
and $T$, the corresponding theoretical value. From the Table
\ref{table1}, we can see, except the error of the neutron, that
all errors given by the CQCM are less than $9\%$, which is much
better than those given by the CVQM. Besides, adding the
theoretical errors in quadrature we get $\sigma_{CQCM}^2=0.046$,
so small than $\sigma_{CVQM}^2=0.12$.

\begin{table}
\begin{center}
\begin{tabular}{c c c c c c c c}
\hline
 & p & n & $\Sigma ^ +  $ & $\Sigma ^ -  $ & $\Xi ^0 $ &
$\Xi ^ -  $
\\
\hline
exp &2.793 & -1.913 & 2.458 & -1.160 & -1.250 & -0.651&-
\\
error&0 & 0&  0.01& 0.025  & 0.014&0.0025&-\\

total error& 0.08 & 0.08 & 0.0806 & 0.0838 & 0.0812 & 0.08&-\\
\hline\hline
CVQM &2.727&-1.818&2.616&-1.021&-1.372&-0.462&$\chi _m ^2$\\
error&-0.0235&-0.0794&0.0641&-0.120&0.0972&-0.290& 16.46 \\
\hline\hline
CQCM &2.745 & -1.705 & 2.668 & -1.118 & -1.261& -0.597&$\chi _m ^2$ \\

error & -0.0173&-0.174&0.0849&-0.0361&0.00950&-0.0837&14.574

\\ \hline\hline
\label{table1}
\end{tabular}
\end{center}
\caption{Magnetic moments (in unit of nucleon magnetic moment) of
the six baryons. The total errors are given by adding to the
experimental error a theoretical error 0.08 in quadrature. The
theoretical results including and not including the sea quark
contributions are listed in the  lines of CQCM and CVQM
respectively. } \label{table1}
\end{table}

The early EMC experiment results of quark spin contributions to
the proton are very rough. The missing spin in this experiment may
come from the gluon polarizations, or orbital angular momentum of
quarks and gluon. But how large they are is still an open
question. In the non-relative quark model, there is no room for
gluons. In the CQCM model with only $12\%$ of $uudd\bar d$ in the
proton, the spin contributions from quark spin and orbital angular
moment are $\Delta u={\textstyle{4 \over 3}}P_3=1.173$, $\Delta
d=-{\textstyle{1 \over 3}}$, and $\Delta l={\textstyle{4 \over
3}}P_{5d}=0.16$. Their sum is equal to 1, which is guarantied by
the wave function Eq.\eqref{eq:wave} to give the total spin of
proton. And we see that the orbital angular moment contributes a
lot.
\section{DISCUSSIONS AND CONCLUSION}

To find the effective degrees of freedom is at the first stage for
studying baryon's structure. The above numerical results have
indicated that the non-perturbative effects of strangeness
component in the nucleon are small. The strangeness content of the
nucleon is purely a sea quark effect and therefor is a clean and
important window to look into the nucleon internal structure and
dynamics. The magnetic moment contribution of strange quark to
nucleon is equal to the measured strange form factor at $Q^2=0$.
The empirical value of strange form factor $ G_M^s (Q^2 = 0.1) =
0.37 \pm 0.20 \pm 0.26 \pm 0.15$ can not give a compelling
evidence for nonzero strange quark effects of proton owing to the
wide uncertainties\cite{Spayde}. The experiment at Mainz with
result $G_E^s + 0.106G_M^s = 0.071 \pm 0.036$ at $Q^2 =
0.108(GeV/c)^2$ \cite{Mass} and the G0 experiment at Jefferson Lab
\cite{Armstrong} may indicate nonzero $ G_E^s$ and $ G_M^s$.
Because these experiments  are carried out all in some special
$Q^2$,the results are still very ambiguous when exploited to
$Q^2=0$. For the strange electric form factor of the proton, the
recent empirical value is $ G_E^s(Q^2=0.1) = - 0.038 \pm
0.042_{(stat)} \pm 0.010_{(syst)}$, which is really consistent
with zero\cite{eform}. A recent lattice result show that the
strange electric form factor is $ G_E^s(Q^2=0.1) = - 0.009 \pm
0.005 \pm 0.003\pm 0.027$ \cite{Leinweber}. These may indicate
that the pairs of strange and anti-strange quarks popping out of
the sea cancel each other so effectively, that they have almost
zero contributions to the proton's magnetic moment, charge, or
mass. This agrees with our numerical results. A very recent
analysis of the complete world set of parity-violating electron
scattering data also gives a result that the strange form factors
are consistent with zero. Further more, recent experiments show
that the strangeness contribution to the proton spin is very small
\cite{smallspin}.

Besides, we like to note that there is another difference between
the CQCM and the $K^+\Lambda$ meson cloud model, apart form the
sign of the strangeness magnetic moment of proton. In the $
K^+\Lambda $ model, the s quark normalized to the probability
$P_{K^ + \Lambda }$ of the $K^ + \Lambda$ configuration yields a
fractional contribution $\Delta S_s= - {\textstyle{1 \over
3}}P_{K^ + \Lambda } $ to the proton spin. Although the  $K^+$,
composed of $u\bar s$ quarks in valence quark model, is
unpolarized, other mechanism may yield $\bar s$ quark polarized
parallel to the initial proton spin\cite{burkardt}. These results
contradict the colored quark cluster model in which only the $\bar
s$ quark give a negative contribution to the proton spin, while
the s quark is unpolarized because the spin symmetry of subsystem
is $[22]_S$. Unfortunately, it is unable to measure the
polarization of single quark experimentally nowadays. More deeply
theoretical analyses are needed.

In the end, the observed non-perturbative quark sea in the nucleon
leads us to reexamine the low lying baryon magnetic moments. We
have deduced these magnetic moments in the CQCM and have give a
discussion of the possible strange component in the nucleon. We
see that the origin of the anomalous moments of quarks discussed
in many works \cite{anomalous} may be from the sea quark
contributions, that the probability of strange component $uuds\bar
s$ in the proton is less than $10\%$ and the probability of
$uudd\bar d$ less than $14\%$, and that the orbital motion has
very significant effects on the spin and magnetic moment in the
non-relative CQCM. Whether there has polarization and about its
sign of strange quark in the proton is still under debate. Our
numerical results favor the non polarization of strange sea quark
in the proton.

We hope future experiments will give us more clues of sea quarks
in these baryons.

\begin{acknowledgments}
 We are grateful to Prof. B. S. Zou for useful discussion.

This work was supported by Chinese Academy of Sciences Knowledge
Innovation Project (KJCX2-SW-No16;KJCX2-SW-No2), National Natural
Science Foundation of China(10435080;10575123).

\end{acknowledgments}

\end{document}